\documentclass[twocolumn,english,floatfixprl,noshowpacs]{revtex4}
\usepackage[T1]{fontenc}
\usepackage[latin9]{inputenc}
\usepackage{float}
\usepackage{graphicx}
\usepackage{esint}

\makeatletter

\@ifundefined{textcolor}{}
{%
 \definecolor{BLACK}{gray}{0}
 \definecolor{WHITE}{gray}{1}
 \definecolor{RED}{rgb}{1,0,0}
 \definecolor{GREEN}{rgb}{0,1,0}
 \definecolor{BLUE}{rgb}{0,0,1}
 \definecolor{CYAN}{cmyk}{1,0,0,0}
 \definecolor{MAGENTA}{cmyk}{0,1,0,0}
 \definecolor{YELLOW}{cmyk}{0,0,1,0}
 }

\usepackage{amsfonts}

\makeatother

\usepackage{babel}

\makeatother

\usepackage{babel}

\begin{document}

\title{Fingerprints of intrinsic phase separation: \\
 magnetically doped two-dimensional electron gas}

\author{H. Terletska and V.~Dobrosavljevi\'{c}}

\affiliation{Department of Physics and National High Magnetic Field Laboratory,
Florida State University, Tallahassee, FL 32310}
\begin{abstract}
In addition to Anderson and Mott localization, intrinsic phase separation
has long been advocated as the third fundamental mechanism controlling
the doping-driven metal-insulator transitions. In electronic system,
where charge neutrality precludes global phase separation, it may
lead to various inhomogeneous states and dramatically affect transport.
Here we theoretically predict the precise experimental signatures
of such phase-separation-driven metal-insulator transitions. We show
that anomalous transport is expected in an intermediate regime around
the transition, displaying very strong temperature and magnetic field
dependence, but very weak density dependence. Our predictions find
striking agreement with recent experiments on $Mn$-doped $CdTe$
quantum wells, a system where we identify the microscopic origin for
intrinsic phase separation. 
\end{abstract}
\maketitle
Two basic routes to the metal-insulator transition have long been
known: the Anderson \cite{anderson58} (disorder-driven) and the Mott
\cite{mott-book90} (interaction-driven) mechanism for electron localization.
In many novel materials, however, the intermediate ``bad metal'' regime
is reached by lightly doping a parent insulator. Here, a new possibility
emerges - the tendency of the few available charge carriers to {}``condense''
into a Fermi liquid - leading to phase separation \cite{gorkov-JETP87}.
However, charge neutrality precludes global phase separation, and
the ground state becomes inhomogeneous \cite{kivelson-rmp03} - even
in the absence of impurities or disorder. Such inhomogeneities can
assume various patterns \cite{dagotto-2005-309}, including bubbles,
stripes, or checkerboard states, and may also cause glassy behavior
even in the limit of very weak disorder \cite{schmalian-prl00,panagopoulos-vlad04prb}.
The relevance of nano-scale phase separation, in its many proposed
forms, has been advocated for the bad metal regimes of cuprates \cite{gorkov-JETP87,kivelson-rmp03}
and manganites \cite{dagotto-book}, and even low density two-dimensional
electron gases (2DEG) \cite{spivak-prl05}. Still, in most of these
cases, no clear and conclusive evidence has been presented that intrinsic
phase separation (IPS) - as driving force for the metal-insulator
crossover - dominates the bad metal regime. While good evidence exists
that here the carrier concentration is indeed inhomogeneous, this
could very well be caused by impurity effects.

To make progress in understanding the relevance of IPS, it would be
very useful and important to identify a model system, that would satisfy
the following criteria: (1) the microscopic origin %
\begin{figure}[h]
\includegraphics[width=2.5in]{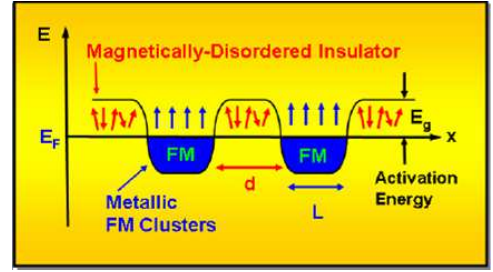}
\caption{(Color online) In presence of magnetically-driven IPS, low-density
carriers are trapped in ferromagnetic {}``bubbles\textquotedblright{}
from which they can escape only by thermal activation. Such activation-limited
transport results in resistivity which is a strong function of temperature
and magnetic field, but a very weak density dependence - the smoking
gun for IPS.\vspace*{-18pt}
}

\end{figure}
for the tendency of phase separation can be established; (2) the system
is clean enough to avoid extrinsic impurity effects; and (3) transport
signatures can be identified that make it possible to distinguish
IPS from spurious disorder effects. In this letter we concentrate
on a specific physical system -- magnetically doped 2DEG -- where
all these criteria can be satisfied. We formulate a simple theoretical
model that describes IPS in this system, and present theoretical results
predicting its transport properties in the metal-insulator crossover
regime.

\emph{Magnetically doped 2DEG: ideal IPS model system. }The criteria
for observing IPS are not easy to satisfy in most available systems.
In many cases, such as for the important family of doped cuprates
\cite{kivelson-rmp03}, its microscopic origin remains highly controversial.
In addition, chemical doping is typically linked with appreciable
amount of disorder created by dopant ions; such is the situation in
most manganite oxides and diluted magnetic semiconductors (DMS). Very
recent experimental work \cite{jan07prb}, however, has presented
striking results on a model system, which seems to be an ideal setting
for observing IPS. Here, a standard 2DEG (formed in\emph{ CdTe} quantum
wells) has been doped by Manganese - a situation that can be described
as a two-dimensional DMS system. Because this 2DEG is created by (iodine)
\emph{modulation-doping}, the impurity ions are far from the mobile
electrons, which dramatically reduces the disorder effects as compared
to the well-known \emph{CdMnTe }bulk (three dimensional) materials.
As a result, the metallic behavior here persists down to remarkably
low carrier densities, even under significant \emph{Mn }doping. In
addition, in this materials (as opposed to the more familiar \emph{GaMnAs}),
\emph{Mn} is neither a donor nor an acceptor.

In this way, devices have been fabricated where, around the critical
carrier density $n_{c}$, there are only a few percents of carriers
per \emph{Mn} ion. The possibility of reaching this regime is very
significant, because it is precisely where one expects the \emph{Mn}-induced
magnetism to favor phase separation \cite{dagotto-book}. Indeed,
both experimental and theoretical considerations have established
\cite{jan07prb} that \emph{CdMnTe }is, under sufficient carrier doping,
a typical Zener (double-exchange) ferromagnet \cite{zener,andersonhasegawa}.
In such materials sufficient doping induces ferromagnetic (FM) correlations
between the \emph{Mn} spins, which can prevail over the usual \emph{Mn-Mn
}antiferromagnetic (AFM) superexchange that exists even in absence
of carriers. At very low doping levels, however, the situation is
more subtle. Here the \emph{average }kinetic energy gain obtained
by FM ordering is not sufficient to balance the energy cost of AFM
superexchange that opposes it. Global FM is then not possible, but
a compromise is reached through phase separation. FM ordering then
prevails only in carrier-rich regions, where phase separation makes
it possible to \emph{locally }achieve a sufficient carrier concentration
favoring Zener ferromagnetism.

\emph{Model. }While the 2D \emph{CdMnTe }system offers an especially
attractive venue to experimentally observe it, magnetically-induced
IPS in a clean low density electron gas is a very robust phenomenon
\cite{dagotto-book}, not confined to two-dimensional systems or the
specific material properties of \emph{CdMnTe. }To emphasize its universal
properties we concentrate on the simplest model that illustrates qualitative
aspects of such an IPS, and is given by the Hamiltonian \begin{equation}
H=H_{mag}+H_{Coul}+H_{imp}.\label{eq:htot}\end{equation}
 The first term here describes a (random) lattice of classical spins
$\vec{S_{i}}$ (representing the Manganese local moments), interacting
with conduction electrons through {}``s-d'' exchange interactions
$J_{sd}$, and with each other through (AFM) superexchange interactions
$J_{ex}$. Following Anderson and Hasegawa \cite{andersonhasegawa},
for simplicity we focus on the $J_{sd}\rightarrow\infty$ limit, where
the conduction electron spins completely align with the local moments

\begin{equation}
H_{mag}=-\sum_{ij}tcos\left(\theta_{ij}/2\right)(c_{i}^{\dagger}c_{j}+c_{j}^{\dagger}c_{i})+J_{ex}\sum_{<ij>}\vec{S_{i}}\vec{S_{j}.}\label{eq:hmag}\end{equation}
 Here, $\theta_{ij}=\arccos(\vec{S_{i}}\vec{S_{j}})$ represents the
relative angle between the neighboring Manganese spins, and $c_{i}^{\dagger}$
is the conduction electron creation operator corresponding to lattice
site $i$. The second term $H_{Coul}$ represents the long-range Coulomb
repulsion between conduction electrons; its principal role is to prevent
global phase separation in order to keep the charge neutrality of
the system. The last term $H_{imp}$ describes (weak) disorder scattering,
which in our system takes the form very similar as in other 2DEG materials
not containing manganese spins.

\begin{figure}[h]
\includegraphics[width=3in]{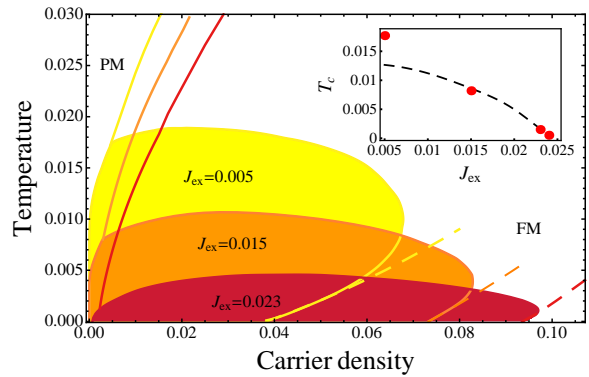}
\caption{(Color online) DMFT phase diagram. Shaded areas show the IPS regions
at three different values of $J_{ex}$. As $J_{ex}$ increases, the
critical density $n_{c}$ increases, while the critical-end-point
temperature $T_{c}$ decreases, until IPS is suppressed at $J>J_{ex}^{c}\approx0.025D$;
the inset shows the analytical (dashed line) low-temperature approximation
(see text) for $T_{c}(J_{ex})$, as compared to the numerical solution
(dots). Note the {}``Pomeranchuk effect'' - the re-entrance of the
coexistence region, due to higher spin entropy of the carrier-poor
PM phase.\vspace*{-12pt}
}

\end{figure}

\emph{IPS at T=0. }To reveal the physical mechanism for IPS in our
model, and understand the respective roles of the two magnetic interactions,
we start our analysis by first constructing the ground state phase
diagram for $H_{mag}.$ In absence of superexchange ($J_{ex}=0$),
the ground state for any doping corresponds \cite{andersonhasegawa}
to complete ferromagnetic (FM) polarization of all $Mn$ spins, as
this allows for maximum kinetic energy of conduction electrons. For
$J_{ex}>0,$ however, at infinitesimal carrier doping the corresponding
energy gain is not sufficient to overcome the cost of superexchange
energy opposing spin alignment. No ferromagnetism is then possible,
instead, the system undergoes phase separation between a FM carrier-rich
phase, and a paramagnetic (PM) %
\footnote{A more detailed analysis shows that our main conclusions are not significantly
modified if one considers AF or spin glass ordering instead of PM
solution we focus on.%
} carrier-poor (carrier-free at $T=0$) phase. For our classical-spin
model, the condition for IPS (phase coexistence) is obtained by equating
the (grand) thermodynamic potentials of the respective phases, viz
$\Omega_{FM}=\Omega_{PM}$. In the FM phase\begin{equation}
\Omega_{FM}=\int_{-D}^{\mu}\rho(w)wdw+J_{ex}/2-\mu\int_{-D}^{\mu}\rho(w)dw,\label{eq:zeroT}\end{equation}
 where $\rho(w)$ is the density of states of (spin polarized) conduction
electrons, and $D\sim t$ is the corresponding half-bandwidth, while
$\Omega_{PM}=0$ %
\footnote{We ignored any spin correlations in the PM phase%
}. We find that IPS emerges when the carrier density $n$ (per $Mn$
ion) satisfies the condition $n<n_{c}$, with \begin{equation}
n_{c}\sim J_{ex}^{d/(d+2)},\label{eq:nc}\end{equation}
 and $\mu_{c}+D\sim J_{ex}^{d/(d+2)}.$ In the entire IPS regime,
both the local carrier concentration in the FM phase $n_{FM}=n_{c}$,
and the corresponding chemical potential $\mu_{FM=}\mu_{c}$ remain
pinned at their critical value (independent of the average density
$n$). As a result, the activation energy gap to the (PM) conduction
band %
\footnote{Due to random-spin orientation, the bandwidth is narrower in the PM
phase; the corresponding band edge $E_{c}\approx-aD$, where $a\approx<\cos(\theta_{ij})>\approx0.71$
. %
} $E_{g}=E_{c}-\mu$ also remains $n$-independent. This simple mechanism,
common to any example of IPS, is the essential cause for very weak
density dependence of (activated) transport in this regime.

\emph{Finite temperature phase diagram - DMFT solution.} At finite
temperatures, the situation is more complicated. First, a fraction
of electrons are transferred from a carrier-rich FM domains into into
the carrier-poor PM regions. Second, at finite temperatures the entropy
of spin fluctuations strongly favors the PM phase. To quantitative
account for both of these effects, we now construct a complete phase
diagram for our system using dynamical mean-field theory (DMFT). This
approach, which is formally exact in the limit of large coordination,
has recently been shown \cite{dagotto-book} to provide quantitatively
accurate results even in low dimensions, for the class of models we
consider. To determine the domain of magnetically-driven IPS, we first
focus on effects described by $H_{mag}$; we will later return to
the role of $H_{Coul}$ and its consequences on transport.

Within DMFT, the lattice model is reduced to solving a single-site
quantum impurity model supplemented by an appropriate self-consistency
condition. For our Anderson-Hasegawa model, the local effective action
takes the form\begin{eqnarray}
S_{eff}(i,\vec{S_{i}}) & = & \int_{0}^{\beta}d\tau\int_{0}^{\beta}d\tau'c_{i}^{\dagger}(\tau)[\delta(\tau-\tau')(\partial_{\tau}+\mu)\nonumber \\
 & + & \Delta_{i}(\tau,\tau',\vec{S})]c_{i}(\tau')-\beta J_{ex}m\vec{S_{i}},\label{eq:seff}\end{eqnarray}
 Here, we have used functional integration over Grassmann fields $c_{i}(\tau)$
describing electron at site $i$. The form of the self-consistency
condition defining the {}``cavity field'' $\Delta_{i}(\tau,\tau',\vec{S})$
depends on the electronic dispersion \cite{georgesrmp}. Within the
DMFT, however, the results do not qualitatively depend on the precise
form of the band structure or the dimensionality; without loss of
generality we therefore use a simple semi-circular model bare density
of states $\rho_{0}(\omega)=\frac{1}{D}\sqrt{D^{2}-\omega^{2}}$.
Using standard methods \cite{georgesrmp}, we find\begin{eqnarray}
\Delta_{i}(w_{n},\vec{S}) & = & (t^{2}/2)(G_{av}^{(i)}(w_{n})+\vec{G}_{m}^{(i)}(w)\vec{\cdot S}),\label{eq:cavity}\end{eqnarray}

\noindent where $G_{av}^{(i)}(w_{n})=<c^{\dagger}(w_{n})c(w_{n})>_{S_{eff}}$
and $\vec{G}_{m}^{(i)}(w_{n})=<\vec{S}c^{\dagger}(w_{n})c(w_{m})>_{S_{eff}}$
are the local Green's functions, and $m=<\vec{S}>_{S_{eff}}$ is the
magnetization. For simplicity, we also use Ising ($S=\pm1)$, as opposed
to Heisenberg spins; we checked that within DMFT this only leads to
very small quantitative changes. These DMFT equations are easy to
solve numerically as a function of the carrier density and temperature,
for several values of the superexchange $J_{ex}$; the resulting phase
diagram is presented in Fig. 2.

\emph{Temperature-driven IPS - Pomeranchuk effect.} Our numerical
solution reproduces Eq. \ref{eq:zeroT} at $T=0$, but interesting
re-entrance (Fig. 2) is found at $T>0$. Physically, the temperature
increase favors phase separation, because of the larger spin entropy
$S_{spin}\sim k_{B}ln2$ associated with the carrier-poor PM phase,
very similar to the effect Pomeranchuk predicted for $He^{3}$ solidification.
Very similar Pomeranchuk-like behavior of phase separation behavior
has been discussed for other systems \cite{spivak-prl05}, and generally
it may give rise to pronounced resistivity maxima just on the metallic
side of the MIT. To clarify this re-entrance, we note that the leading
low-temperature behavior can be analytically obtained from Eqs. (\ref{eq:seff},\ref{eq:cavity}),
by performing an appropriate Sommerfeld expansion. To leading order
$\Omega_{fm}(T)=\Omega_{fm}(T=0)+O(T^{2}),$ while $\Omega_{pm}(T)=-k_{B}T\ln2+O(T^{2})$,
giving an expression for the high-density boundary of the coexistence
dome (dashed line on Fig. 2) of the form\begin{equation}
T_{coex}^{+}=\frac{J_{ex}}{2k_{B}\ln2}\left(\left(n/n_{c}\right)^{(d+2)/d}-1\right).\label{eq:lowT}\end{equation}
 A similar argument gives an expression for critical temperature of
IPS \begin{equation}
T_{c}(J_{ex})\approx C(J_{ex}^{c}-J_{ex}),\label{eq:tc}\end{equation}
 where for our model $J_{ex}^{c}\approx0.025D$, and $C\approx0.88$.
This shows, in perfect agreement with our numerics (see inset of Fig.
2), that while $n_{c}$ increases with $J_{ex}$ (see Eq. \ref{eq:nc}),
the critical temperature $T_{c}$ decreases, until phase separation
is suppressed for $J_{ex}>J_{ex}^{c}$.

\noindent %
\begin{figure}[H]
\includegraphics[width=3.17in]{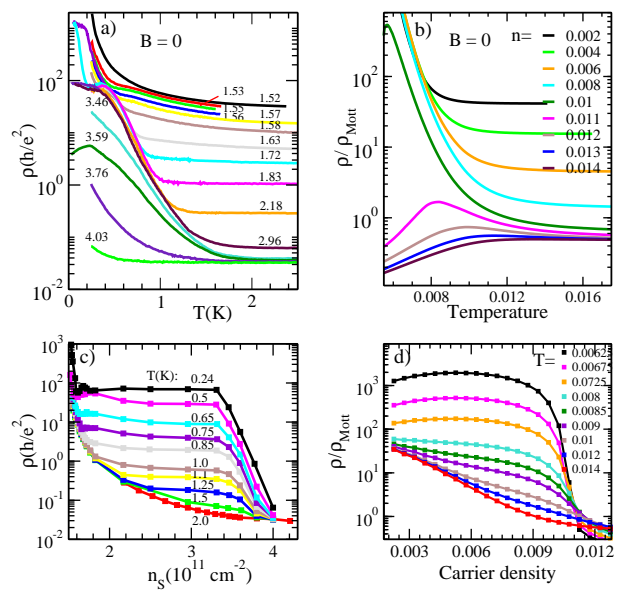}
\caption{(Color online) (a) and (b) show the resistivity as function of temperature at different
carrier densities. For comparison, we contrast the experimental results
\cite{jan07prb} (left panel) and our DMFT predictions (right panel).
A sharp upturn in resistivity in the low temperature regime, and its
weak density dependence are direct consequences of IPS. Plots (c)
and (d) show the same data as function of carrier density at fixed
temperatures to emphasize weak density dependence of resistivity below
the percolation threshold. It also reflects the fact that within this
region activation energy as well as the size of the PS regions is
density independent (see text).\vspace*{-12pt}
}

\end{figure}

\emph{Phase separation and transport.} As density increases in the
IPS region, the volume fraction $x$ of the carrier-rich FM phase
determined from $n=xn_{FM}+(1-x)n_{PM}$ also increases, until a percolation
concentration is reached and metallic transport sets in. In presence
of long-range Coulomb interactions, the characteristic size of FM)
carrier-rich domains is limited by the charging energy, which can
be easily estimated.

However, within a commonly-used random-resistor picture of transport
(which is well known to be adequate for the kind of system we consider)
the critical behavior is largely insensitive to the individual domain
size - with the only relevant parameter being the relative volume
fraction $x$. Thus, to obtain a quantitative estimate of transport
properties for our model, it suffices to calculate the local temperature-dependent
resistivity for each of the two phases $\rho_{FM/PM}(n,T)$. This
task is easily carried out using standard methods \cite{Letfulov,georgesrmp}
within DMFT model allowing us to determine the resistivity from the
solution of our DMFT equations. This calculation accounts for density
and temperature dependent resistivity $\rho_{FM/PM}(n,T)$ due to
scattering of conduction electrons off Manganese spins in the respective
phases.

To make a meaningful comparison with experiments, however, we also
have to account for conventional impurity scattering while considering
the total resistivity behavior. This weak impurity scattering in our
system which, by the way, is unrelated to the physics of IPS and is
more pronounced at higher temperatures, is of the same form as in
conventional 2DEG systems \cite{AndoRevModPhys.54.437}. In particular,
as can be seen from the high-temperature transport data of Ref. \cite{jan07prb},
it has very weak temperature dependence but an appreciable density
dependence, which we obtain by fitting the experimental data, giving
$\rho_{imp}\approx\rho_{o}\exp\{-An\}+C$, (here we used $\rho_{o}=3600\rho_{Mott}$,
$A=850$ and $C=0.2\rho_{Mott}$; resistivity is in units of $\rho_{Mott}$
\footnote{$\rho_{Mott}$ is the Mott limit of the maximum metallic resistivity;
to determine it we used $E_{F}\tau=1$ ($E_{F}$ is Fermi energy and
$\tau$ is scattering time); N. E. Hossein \textit{et al.}, Philosophical
Magazine \textbf{84,} 2847 (2004). %
}). 

The net resistivity within the IPS region is then calculated by solving
the random resistor problem, where we use the familiar effective-medium
approximation (with $\rho_{med}$ being determined as in (Eq. 5.7)
of \cite{kirkpatrick-rmp94}), giving \[
\rho(n,T)=\rho_{med}(\rho_{fm},\rho_{pm})+\rho_{imp}.\]

The resulting resistivity curves are given in Fig. 3; as we can see,
our model seems to capture all the qualitative features of the experimental
data. Below the (temperature-dependent) percolation threshold, where
we see a sharp increase in resistivity, the transport is activated
with essentially density-independent activation energy. At higher
densities, we note the well-pronounced resistivity maximum at low
temperatures, reflecting the re-entrant Pomeranchuk-like IPS behavior
we described above.

\noindent \emph{$\quad$CMR behavior}. As in other double-exchange
ferromagnets, a moderate magnetic field reduces the IPS region, quickly
lowers the relevant activation energy, and leads to colossal magnet-resistance
(CMR) behavior (Fig. 4). 

\noindent %
\begin{figure}[H]
 \includegraphics[width=1.62in]{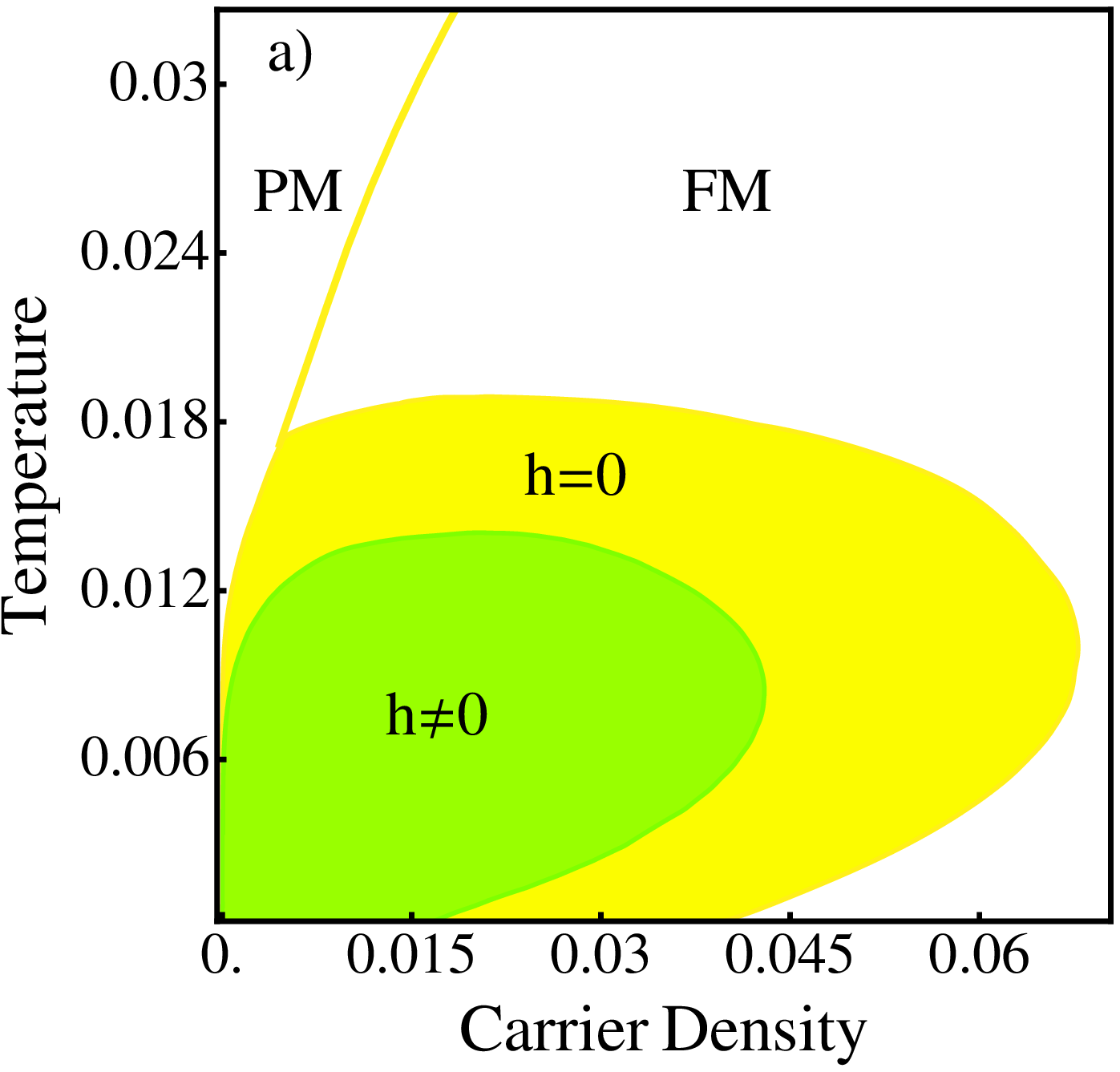}\includegraphics[width=1.6in]{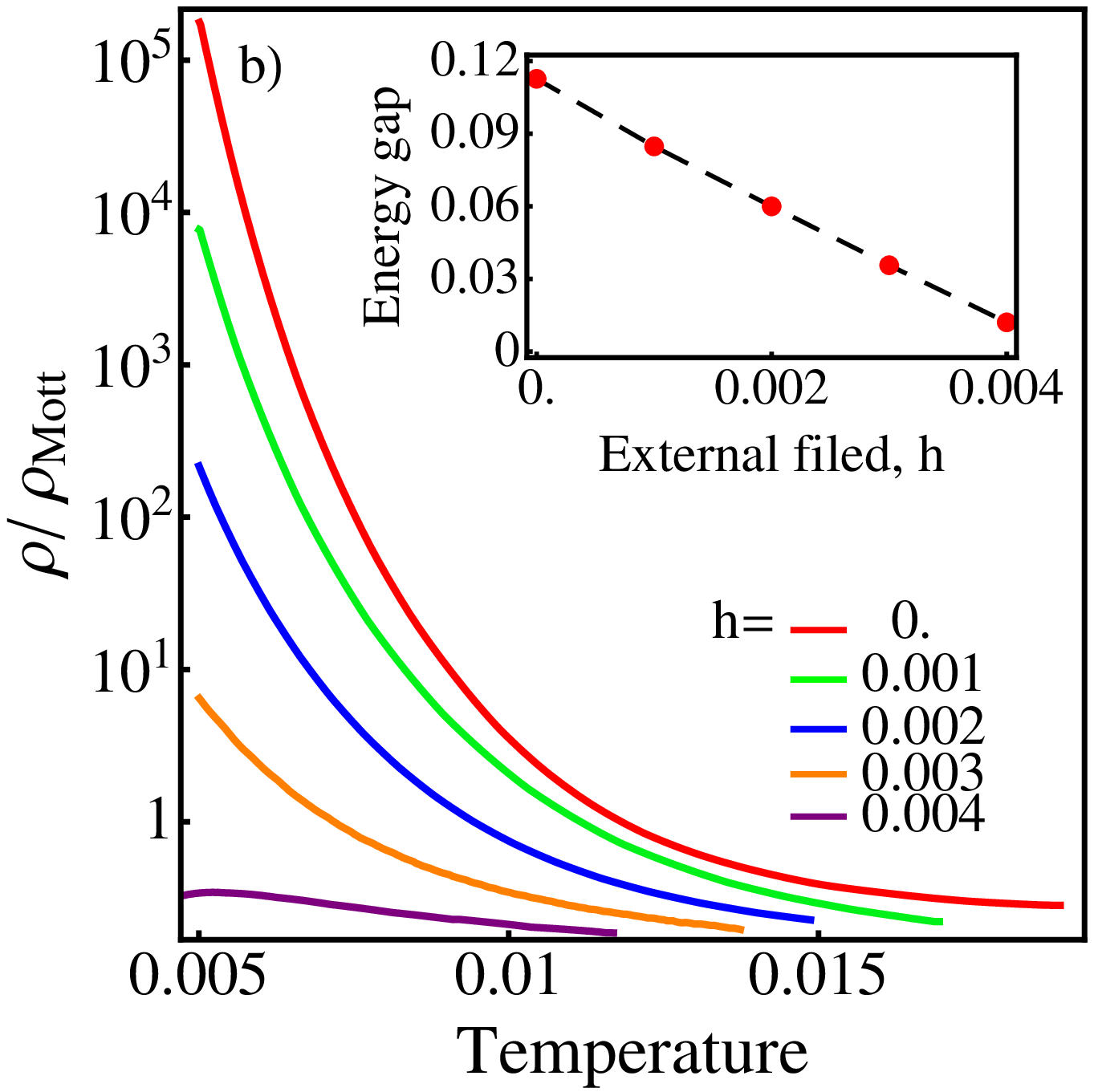}
\caption{(Color online) Phase diagram (a) and transport (b) in presence of an external magnetic
field $h$. Panel (a) shows how the temperature and density parameter
range of IPS is quickly reduced in nonzero magnetic field, and (b)
depicts the negative magnetoresistance within this IPS region (also
seen in \cite{jan07prb}). The inset of (b) shows the decrease of
activation energy $E_{g}$ (which was extrapolated from the resistivity
vs. temperature plots) with magnetic field. \vspace*{-12pt}
}

\end{figure}

\noindent However, this behavior is restricted to the coexistence
region, and as such can serve as an indication for IPS behavior-again
in close agreement with experiments\cite{jan07prb}. 

\emph{Conclusions.} In this letter we presented and solved a simple
microscopic model for magnetically-driven IPS behavior. Given the
simplicity of our theoretical model, the agreement with all the qualitative
features of the experimental data is remarkable. Most importantly,
we observe how, within the IPS regime, all the resistivity curves
{}``merge'' together below a temperature scale set by $T_{c}$ for
phase separation. This striking feature simply reflects the fact that
the chemical potential - setting the activation scale in this problem
- is fixed and density-independent everywhere under the IPS dome.
Such behavior is quantitatively reproduced in our model for magnetically
driven IPS. However, very general thermodynamic principles guarantee
that the very same behavior must be at play in virtually any system
where transport is driven by a tendency to inherent phase separation
- and not the disorder effects. In this work we identified the first
physical system which presents plausible and convincing evidence for
the IPS scenario. From a more general point of view, our results indicate
how should one search for other examples where this fascinating physics
may be recognized and properly interpreted.

We thank P. Schlottmann and S. von Molnar for fruitful discussion.
This work was supported by the NSF grant DMR-0542026 and the National
High Magnetic Field Laboratory.


\begin{thebibliography}{16}
\bibitem{anderson58} P. W. Anderson, 
Phys. Rev. \textbf{109}, 1492 (1958).

\bibitem{mott-book90}N.~F.Mott, \textit{Metal-Insulator Transition.}
\newblock (Taylor \& Francis, London, 1990).

\bibitem{gorkov-JETP87} L.~P. Gor'kov {\em et al.}, \newblock
JETP Lett. \textbf{46}, 420 (1987).

\bibitem{kivelson-rmp03} 
S.~A. Kivelson {\em et al.}, Rev. Mod. Phys. \textbf{75}, 1201
(2003).

\bibitem{dagotto-2005-309} E. Dagotto, 
\newblock Science \textbf{309}, 257 (2005).

\bibitem{schmalian-prl00} J.~Schmalian {\em et al.}, 
\newblock Phys. Rev. Lett. \textbf{85}, 836 (2000).

\bibitem{panagopoulos-vlad04prb} C.~Panagopoulos {\em et al.},
\newblock Phys. Rev. B, \textbf{72}, 014536 (2005).

\bibitem{dagotto-book} E.~Dagotto, \newblock {\em Nanoscale Phase
Separation and Colossal Magnetoresistance} \newblock (Springer-Verlag,
Berlin, 2002).

\bibitem{spivak-prl05} R.~Jamei{\em et al.}, 
\newblock Phys. Rev. Lett. \textbf{94}, 056805 (2005).

\bibitem{jan07prb} J.~Jaroszy\'{n}ski {\em et al.}, 
\newblock Phys. Rev. B, \textbf{76}, 045322 (2007).

\bibitem{zener} C.~Zener, 
\newblock Phys. Rev. \textbf{82}, 403 (1951).

\bibitem{andersonhasegawa} P.~W. Anderson {\em et al.}, 
\newblock Phys. Rev. \textbf{100}, 675 (1955).

\bibitem{georgesrmp} A.~Georges {\em et al.}, 
\newblock Rev. Mod. Phys. \textbf{68}, 13 (1996).

\bibitem{Letfulov} B.~M. Letfulov {\em et al.}, 
\newblock Phys. Rev. B, \textbf{64}, 174409 (2001).

\bibitem{AndoRevModPhys.54.437} T. Ando {\em et al.}, 
\newblock Rev. Mod. Phys., \textbf{54}, 437 (1982).

\bibitem{kirkpatrick-rmp94} D.~Belitz {\em et al.}, 
\newblock Rev. Mod. Phys. \textbf{66}, 261 (1994).

\end{thebibliography}

\pagebreak{} 
\end{document}